\documentclass{jfm}
\usepackage{graphicx}
\usepackage{epstopdf, epsfig}
\usepackage{color}

\shorttitle{Turbulent Taylor-Couette flow with stationary inner cylinder}
\shortauthor{R. Ostilla-M\'onico, R. Verzicco and D. Lohse}

\title{Turbulent Taylor-Couette flow with stationary inner cylinder }

\author{R. Ostilla-M\'onico\aff{1,2}
  \corresp{\email{rostillamonico@g.harvard.edu}},
  R. Verzicco\aff{1,3}
 \and D. Lohse\aff{1,4}}

\affiliation{
\aff{1}Physics of Fluids Group, Mesa+ Institute and J.M. Burgers Centre for Fluid Dynamics, University of Twente, P.O. Box 217, 7500 AE Enschede, The Netherlands
\aff{2}School of Engineering and Applied Sciences and Kavli Institute for Bionano Science and Technology, Harvard University, Cambridge, MA 02138, USA.
\aff{3}Dipartimento di Ingegneria Industriale, University of Rome ``Tor Vergata", Via del Politecnico 1, Roma 00133, Italy
\aff{4}Max Planck Institute for Dynamics and Self-Organization, 37077 G\"ottingen, Germany
}

\begin{document}

\maketitle

\begin{abstract}

A series of direct numerical simulations of Taylor-Couette (TC) flow, the flow between two coaxial cylinders, with the outer cylinder rotating and the inner one fixed, were performed. Three cases, with outer cylinder Reynolds numbers $Re_o$ of $Re_o=5.5\cdot10^4$, $Re_o=1.1\cdot10^5$ and $Re_o=2.2\cdot10^5$ were considered. The radius ratio $\eta=r_i/r_o$ was fixed to $\eta=0.909$ to mitigate the effects of curvature. Axially periodic boundary conditions were used, with a vertical periodicity aspect ratio $\Gamma$ fixed to $\Gamma=2.09$. Being linearly stable, outer cylinder rotation TC flow is known to have very different behaviour than pure inner cylinder rotation TC flow. Here, we find that the flow nonetheless becomes turbulent, but the torque required to drive the cylinders and level of velocity fluctuations was found to be smaller than those for pure inner cylinder rotation at comparable Reynolds numbers. The mean angular momentum profiles showed a large gradient in the bulk, instead of the constant angular momentum profiles of pure inner cylinder rotation. The near-wall mean and fluctuation velocity profiles were found to coincide only very close to the wall, showing large deviations from both pure inner cylinder rotation profiles and the classic von Karman law of the wall elsewhere. Finally, transport of angular velocity was found to occur mainly through intermittent bursts, and not through wall-attached large-scale structures as is the case for pure inner cylinder rotation.

\end{abstract}


\section{Introduction}

Taylor-Couette (TC) flow, the flow between two coaxial and independently rotating cylinders can present different types of transition to turbulence. For vanishing viscosity, TC flow is linearly unstable if $|r_o\omega_o| < |r_i\omega_i|$, where $r_{i,o}$ are the inner and outer cylinder radii, and $\omega_{i,o}$ their angular velocities \citep{ray17}, due to the centrifugal forces. If viscosity is considered, a minimum rotation strength is required to overcome the viscous damping. For pure inner cylinder rotation, once this stability threshold is crossed, TC flow presents a supercritical transition to turbulence, where the purely azimuthal flow develops large-scale structures which fill up the entire gap and effectively redistribute angular momentum \citep{tay23,far14}. With increasing rotation, these large scale structures undergo a series of transitions from laminar Taylor vortices to wavy Taylor vortices to modulated wavy Taylor vortices to turbulent Taylor vortices \citep{and86}. Even at very large Reynolds numbers these structures have been observed to persist in some regions of the parameter space, i.e.\ at $Re\sim\mathcal{O}(10^5)$ in simulations \citep{ost14d} and then at $Re\sim\mathcal{O}(10^6)$ in experiments \citep{hui14}. The flow field shows a clear lack of statistical spatial homogeneity and these ``rolls'' cause substantially large deviations of the near-wall velocity profiles of TC flow from the classic wall-turbulence profiles of channels and pipes. This effect has been attributed to the role of curvature and the centrifugal instability \citep{ost16}. For an overview of supercritical TC flow at large Reynolds numbers, we refer the reader to the review of \cite{gro16}.

If instead, the \emph{outer} cylinder is rotated, and the inner cylinder is kept fixed, the flow undergoes a sub-critical transition to turbulence. This transition is quite different from the supercritical transition detailed previously, as the flow does not go undergo a series of changes from less commonplex to more complex flow patterns, but instead makes a sudden transition to turbulence, either in localized spots or filling the entire gap. {While \cite{tay36} found evidence for this sub-critical transition by measuring the torques and how they deviated from the predictions for steady flow}, this transition was first systematically studied by \cite{col65}, who found that for low outer cylinder Reynolds numbers, $Re_o=dr_o\omega_o/\nu$ with $\nu$ the fluid kinematic viscosity and $d$ the gap-width, $d=r_o-r_i$, intermittent turbulent patches coexisted with laminar flow, with well-defined interfaces. The persistence time of these patches increased with increasing inner cylinder Reynolds number $Re_i=d\omega_ir_i/\nu$, and so did the turbulent fraction, until the flow was fully turbulent. For the lower $Re_o$ range, the flow had to be started in a supercritical state, such that the centrifugal instabilities provided an initial perturbation for the generation of turbulence. For larger values of $Re_o$, the flow no longer required the centrifugal instability to transition to turbulence, and could remain exclusively in the sub-critical region and still see a spontaneous, or ``catastrophic'' transition to turbulence.

Studies of subcritical TC flow continued through the years, both theoretically, in an attempt to develop non-linear stability criteria, as well as numerically and experimentally. The focus of many of these studies was on the sharp turbulent-laminar interface  and on spiral turbulence, a particular flow where the bursts took a spiral shape \citep{vanatt66,and83}. We refer the reader to the thesis of \cite{bor14} for a detailed historical overview of subcritical TC flow studies. Subcritical transitions to turbulence have been well studied in the past, and are an active area of research, as not only TC flow, but also pipe and channel flows present a subcritical transition to turbulence. For a comprehensive overview of this field, we refer the reader to the review by \cite{eck08}. 

A recent systematic study of pure outer cylinder rotation (OCR) in TC flow was performed by \cite{bur12}, who experimentally studied in detail the effect of gap-width and end-plate configurations on the transition to turbulence. These authors also performed velocimetry in the bulk and found that regions of high turbulence were associated to high shear. Earlier, \cite{bor10} had already provided evidence for super-exponential dependence on the Reynolds number of the decay times of turbulence. Therefore, it seems that high Reynolds number outer cylinder TC flow is turbulent for extremely long time scales. This regime has not been well characterized: \cite{bur12} did not provide velocimetry close to the walls. The other experimental studies by \cite{pao11,pao12} only provided torque measurements for pure OCR, which indicated values well above the values for laminar flow but also much lower than the torque values for pure inner cylinder rotation. However, experiments are limited by the necessary presence of end-plates to provide flow confinement, and this could potentially affect the physics. Numerical studies of pure OCR in an infinite (periodic) domain are limited to \cite{deg14}, who considered Reynolds numbers near the transition to turbulence. 

In this manuscript, we conducted a series of direct numerical simulations (DNS) of axially periodic and fully turbulent TC flow with only outer cylinder rotation, in an attempt to isolate and study subcritical behavior of TC flow, and to eliminate the effect of perturbations arising at the end plates. We consider pure outer cylinder rotation as it does not have the complex combination of sub- and supercritical behavior seen in turbulent counter-rotating TC flow \citep{gil12,bra13b,gro16}. The simulated TC geometry is a narrow-gap system, which produces very strong rolls in the pure inner cylinder rotation (ICR) case, and limits the effect of strong curvature, which causes very different flow physics \citep{ost16}. In this manuscript, we extend the analysis of \cite{ost16} in an attempt to understand which pieces of the flow physics come from the centrifugal (in)stability and to reveal and quantify the differences between supercritical and subcritical TC turbulence.  

\section{Simulation details}

The DNS were performed using an energy-conserving second-order centered finite-difference code with fractional time stepping \citep{ver96,poe15}. This code has been extensively used and validated for TC flow. The radius ratio $\eta=r_i/r_o$ was chosen as $\eta=0.909$ as in \cite{ost16}, to mitigate curvature effects. The aspect ratio $\Gamma=L_z/d$, where $L_z$ is the axial periodicity length was taken as $\Gamma=2.09$. To reduce computational costs, a rotational symmetry order $n_{s}=20$ was imposed, which results in a minimum azimuthal extent of $\pi$-gap widths at the inner cylinder. This choice of $n_{sym}$ and $\Gamma$ results in computational boxes which are large enough to show sign changes of the azimuthal velocity autocorrelation functions at the mid-gap, as was already observed in \cite{ost15} for pure inner cylinder rotation. The size of the time-steps was chosen dynamically by imposing that the maximum Courant-Freiderichs-Lewy (CFL) number in the grid is 0.5.

Three different outer cylinder Reynolds numbers were simulated: $Re_o=5.5\cdot10^4$, $Re_o=1.1\cdot10^5$ and $Re_o=2.2\cdot10^5$. These outer cylinder Reynolds numbers have an equivalent shear Reynolds number $Re_s=2| \eta Re_o - Re_i|/(1+\eta)$ \citep{dub05} to pure ICR rotations at $Re_i=5\cdot 10^4$, $Re_i=1\cdot 10^5$, and $Re_i=2\cdot 10^5$. These Reynolds numbers are much larger than the transitional Reynolds numbers for spiral turbulence $Re_o \sim 5000$, \citep{col65,and83,deg14}, and are in the regime where no spiral structures are seen in experiments \citep{bur12}. With the largest Reynolds number, an inner cylinder frictional Reynolds number $Re_{\tau,i}=u_{\tau,i}d/\nu$ of up to $Re_{\tau,i}=1220$ is achieved, where the inner cylinder frictional velocity is defined as $u_{\tau,i}=\sqrt{\tau_w/\nu}$ with $\tau_w$ the shear stress at the cylinder wall. The outer cylinder frictional Reynolds number (velocity) is simply $Re_{\tau,o}=\eta Re_{\tau,i}$ ($u_{\tau,o}=\eta u_{\tau,i}$). For convenience we define the inner cylinder viscous length as $\delta_{\nu,i}=\nu/u_{\tau,i}$, the non-dimensional distance from the wall $\tilde{r}=(r-r_i)/d$ the non-dimensional axial coordinate $\tilde{z}=z/d$, and the non-dimensional angular velocity $\tilde{\omega}=\omega/\omega_o$, with the angular velocity $\omega=u_\theta/r$. 

We also note that the lowest Reynolds number simulated is about one order of magnitude larger than the estimated Reynolds number for transition at $\eta=0.909$ by \cite{bur12}. It was impossible with our simulations to achieve stable turbulent states at Reynolds numbers lower than $Re_o=5.5\cdot10^4$, probably due to the small computational box used. To perform the simulations, we first started a simulation with $Re_o=1.1\cdot10^5$ and a stationary inner cylinder with white noise of $\mathcal{O}(r_o\omega_o)$. After a very long transient of about 1000 large eddy turnover times based on $d/(r_o\omega_o\eta)$, a statistically stationary state was reached. The transients were significantly longer than those of pure ICR TC flow, as $u_\tau$ is a factor two to three times smaller. This state was used as initial condition for both the $Re_o=2.2\cdot10^5$ and $Re_o=5.5\cdot10^4$ simulations, and the mesh was either coarsened or refined to ensure a correct balance between accuracy and speed of computation. Attempting to start simulations at lower $Re_o$ from initial conditions at $Re_o=5.5\cdot 10^4$ resulted in divergence of the fields due to unclear reasons. If white noise of order $\mathcal{O}(r_o\omega_o)$ was used at $Re_o=1.1\cdot 10^4$, the system would slowly relaminarize and return to the purely azimuthal state. 

After the transients, the simulations were run (at least) for an additional 67 large-eddy turnover times based on $\tilde{t}=\eta r_o\omega_ot/d$. The temporal convergence was assured by checking the radial dependence of the angular velocity current $J^\omega$, defined as $J^\omega=r^3(\langle u_r\omega\rangle_{\theta,z,t}-\nu\partial_r\langle\omega\rangle_{\theta,z,t})$ \cite{eck07b}, where $\langle ... \rangle_{x_i}$ denotes averaging with respect to $x_i$. $J^\omega$ should have no radial dependence when averaged for an infinite time, however, for finite time statistics we considered that deviations smaller than 3\% from the average value in the bulk were sufficient, as these were associated to deviations of the time-averaged torque ($J^\omega$ at the cylinders) at both cylinders smaller than $1\%$, a value we have previously used \citep{ost16}. The computational domain was uniformly discretized in the azimuthal and axial directions, while a clipped Chebychev type clustering was used in the radial direction. Full details of the numerical resolution used are provided in Table \ref{table:num}. The table includes pure ICR data (i.e.\ the I1 and I2 cases) from \cite{ost16} (referred to there as R1 and R2) for comparison. The I1 case has the same $Re_\tau$ as the O2 case, while the I2 case has the same driving shear as the O2 case. 

\begin{table}
  \begin{center} 
  \def~{\hphantom{0}}
  \begin{tabular}{|c|c|c|c|c|c|c|c|c|c|c|}
  \hline
 Case & $Re_i$ & $Re_o$ & $N_\theta$ & $N_r$ & $N_z$ & $\Delta r^+$ & $\Delta z^+$ & $r_i\Delta\theta^+$ & $Re_{\tau,i}$ & $Nu_\omega$  \\ 
  \hline
  O0 & 0 & $5.5\cdot10^4$ & $384$ & $768$ &  $512$ & $0.2$-$1.6$ & $3.3$ & $6.6$ & $402$ & $11.6\pm 0.2$ \\
  O1 & 0 & $1.1\cdot10^5$ & $384$ & $1024$ & $768$ & $0.2$-$2.1$ & $3.8$ & $11.5$ & $703$ & $18.3\pm 0.8$ \\
  O2 & 0 & $2.2\cdot10^5$ & $768$ & $1024$ & $1024$ & $0.3$-$3.6$ & $5.0$ & $9.9$ & $1220$ & $26.2\pm 1.3$ \\ 
  I1 & $1\cdot10^5$ & 0 & $1024$ & $1024$ & $2048$ & $0.3$-$4.1$ & $2.7$ & $9.1$ & $1410$ & $69.5\pm0.2$ \\ 
  I2 & $2\cdot10^5$ & 0 & $1536$ & $1536$ & $3072$ & $0.3$-$5.2$ & $3.4$ & $11.4$ & $2660$ & $126\pm2.1$ \\ 
    \hline
 \end{tabular}
 \caption{Details of the numerical simulations. The first column is the name with which the simulation  will  be  refereed  to  in  the  manuscript.  The second and third column are the inner and outer cylinder Reynolds number. The second to fourth columns represent the amount of points in the azimuthal, radial and axial directions. The fifth column represents the minimum and maximum resolution in the radial direction normalized with the inner cylinder wall unit. The sixth and seventh columns show the axial and azimuthal resolutions (at the inner cylinder) in inner cylinder wall units. The eighth column refers to the inner cylinder frictional Reynolds number, and the last column shows the torque non-dimensionalized as a pseudo-Nusselt number. }
 \label{table:num}
\end{center}
\end{table}

\section{Results}

We first focus on the torque to drive the cylinders. At comparable Reynolds numbers, a smaller torque is required for pure OCR than for pure ICR, as can be seen from Table \ref{table:num}. The torque, non-dimensionalized as a pseudo-Nusselt number $Nu_\omega=J^\omega/J^\omega_{pa}$, where $J^\omega_{pa}$ is $J^\omega$ for the purely azimuthal flow, is approximately a factor four smaller. As a direct consequence of this, the frictional Reynolds number $Re_\tau$ is approximately a factor two lower because $Re_\tau \sim \sqrt{Nu_\omega}$. This results in smaller values of $u_\tau$ for pure OCR, and thus the longer transients observed in the DNS.

Transport of angular velocity from across the gap is much more inefficient in the case of subcritical turbulence, something that can be expected from the ``optimal'' transport results of \cite{gil12,pao12}, where the driving torque drastically decreases with the appearance of the radial partitioning into sub-critical and super-critical zones.  However, unlike the quasi-Keplerian case, where $|r_o^2\omega_o| > |r_i^2\omega_i|$ and $|\omega_o|<|\omega_i|$ which were found both numerically \cite{ost14b} and experimentally \cite{nor15} to not sustain angular velocity transport across the gap, in pure OCR there is still turbulence present, and the flow is not purely azimuthal as $Nu_\omega\ne1$. In the pure OCR rotation case, both the gradients of angular velocity and angular momentum point in the same direction, i.e.\ inwards, while in the quasi-Keplerian cases, they point in different directions, i.e.\ inwards for the angular momentum and outwards for the angular velocity. 

\begin{figure}
  \centering
    \includegraphics[width=0.48\textwidth]{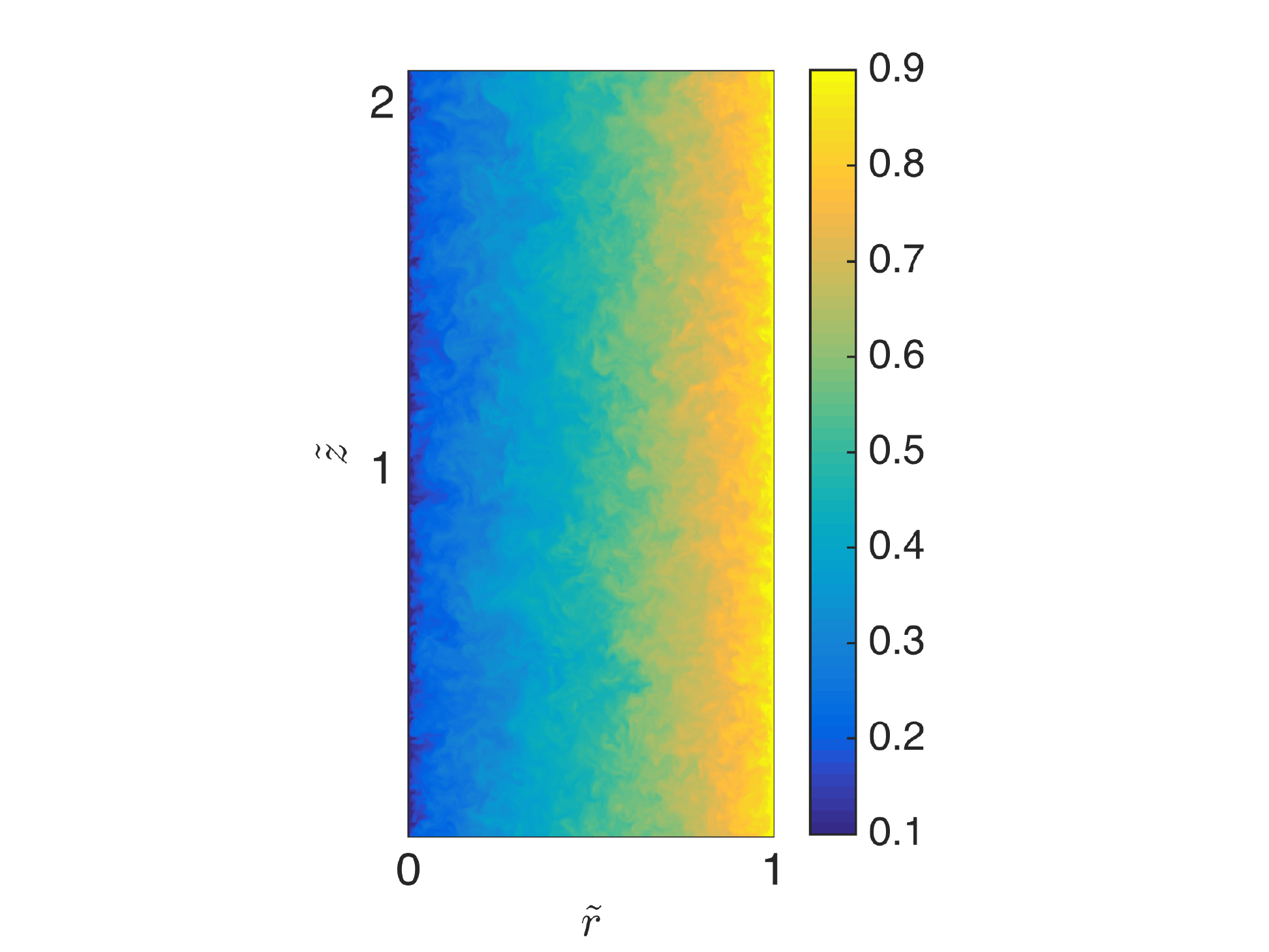}
     \includegraphics[width=0.48\textwidth]{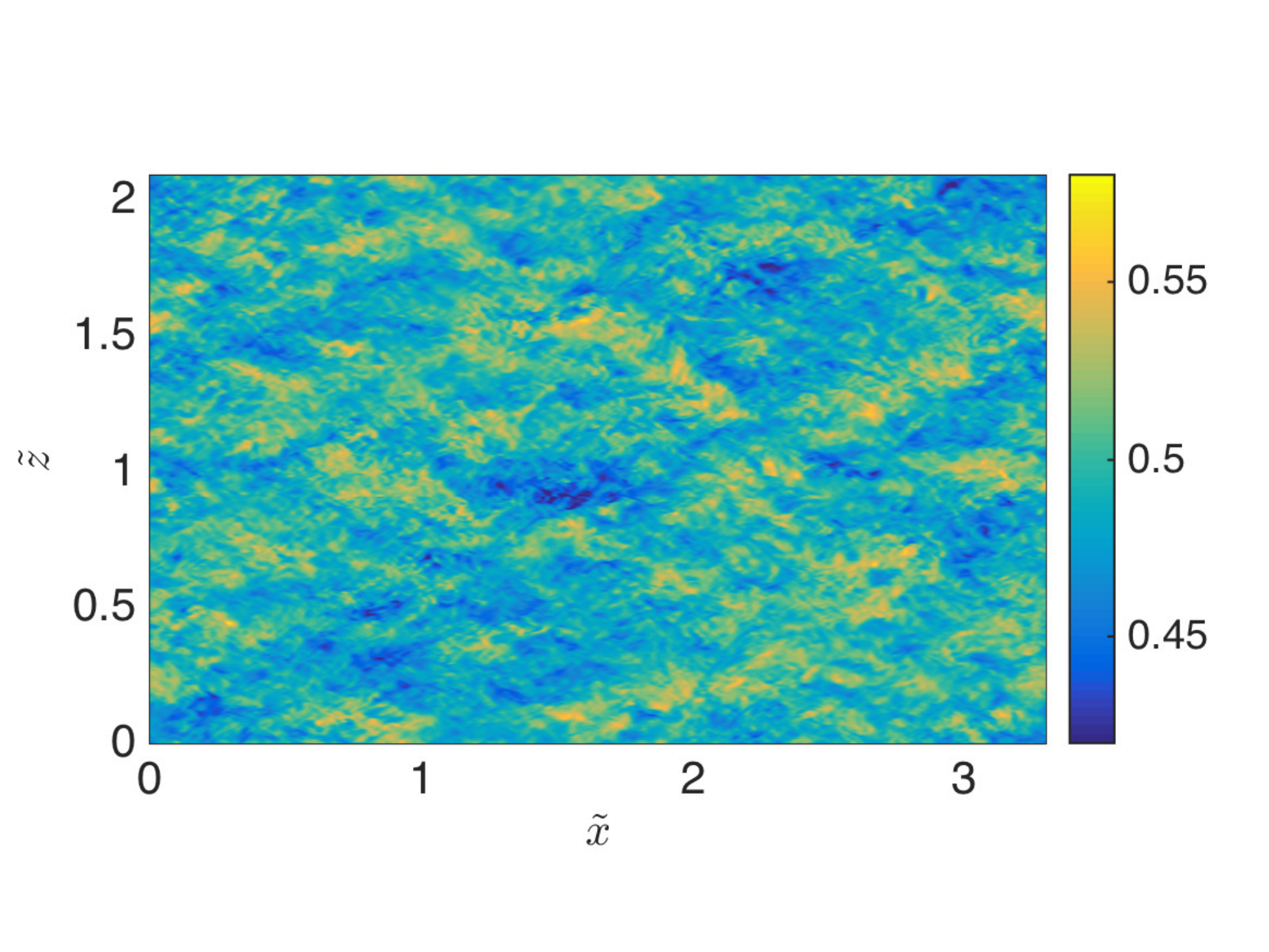}
     \caption{ Pseudocolour plot of the instantaneous angular velocity for the O2 case for a constant azimuth (left) and at the mid-gap (right). No large-scale structures can be seen, as is seen for pure inner cylinder rotation in \cite{ost15,ost16}.}
\label{fig:q1ststs}
\end{figure}

To understand why the torque is lower for pure OCR, we visualize the flow field in figure \ref{fig:q1ststs}, which shows a pseudocolor plot of the instantaneous angular velocity for an azimuthal cut (left) and at the mid-gap (right) for the O2 case. A complete absence of the large-scale rolls can be seen. These figures can be compared to Figs. 1 and 3 of \cite{ost15}, visualizations of the instantaneous velocities for the I1 case. For the same geometrical parameters, and similar Reynolds numbers, the velocities in the pure IRC cases have marked axial inhomogeneities. The existence of rolls has been linked to increased transport \citep{bra13b,gro16}, so from this alone we can expect a smaller $Nu_\omega$.

\begin{figure}
  \centering
    \includegraphics[width=0.48\textwidth]{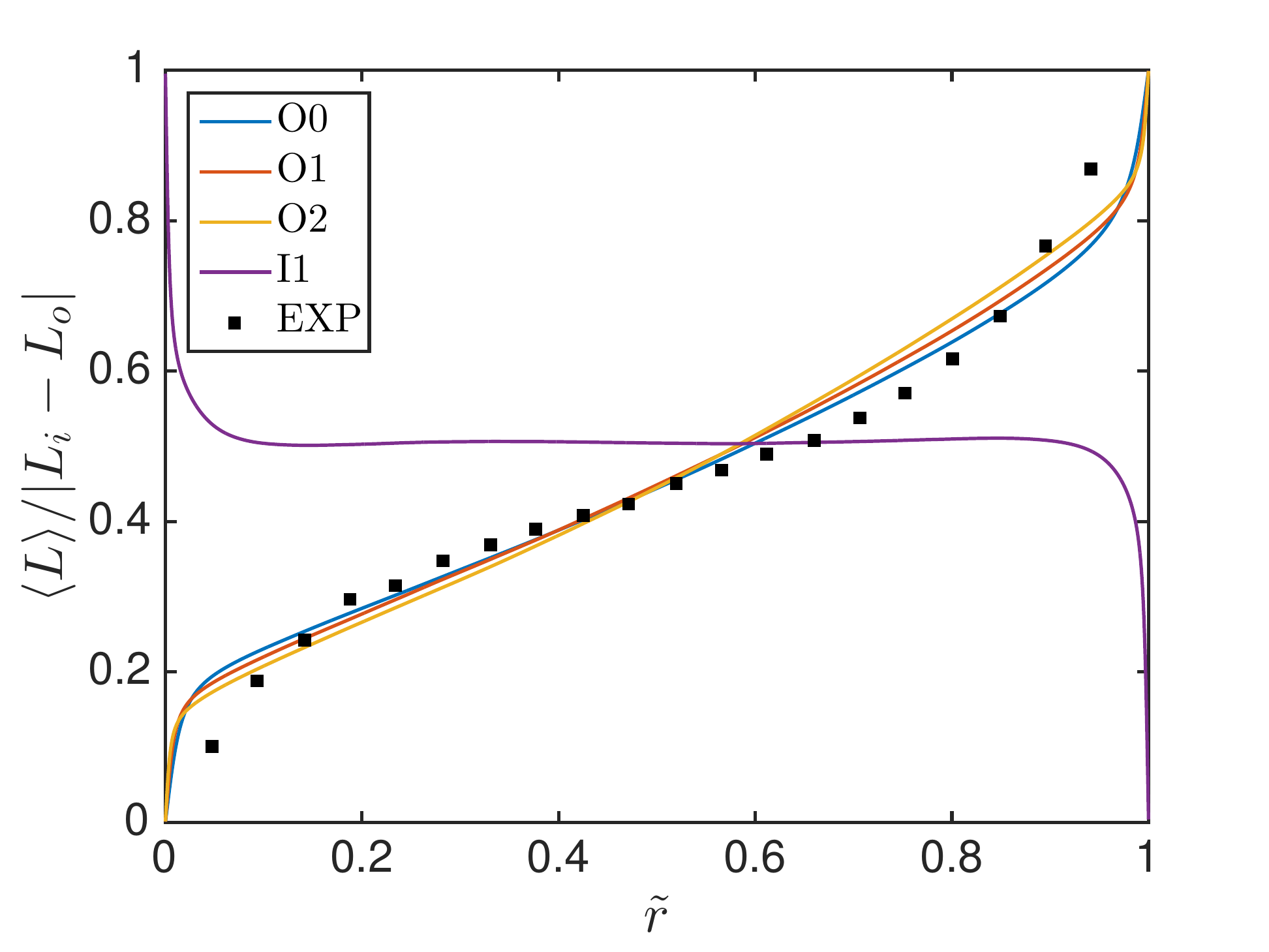}
     \includegraphics[width=0.48\textwidth]{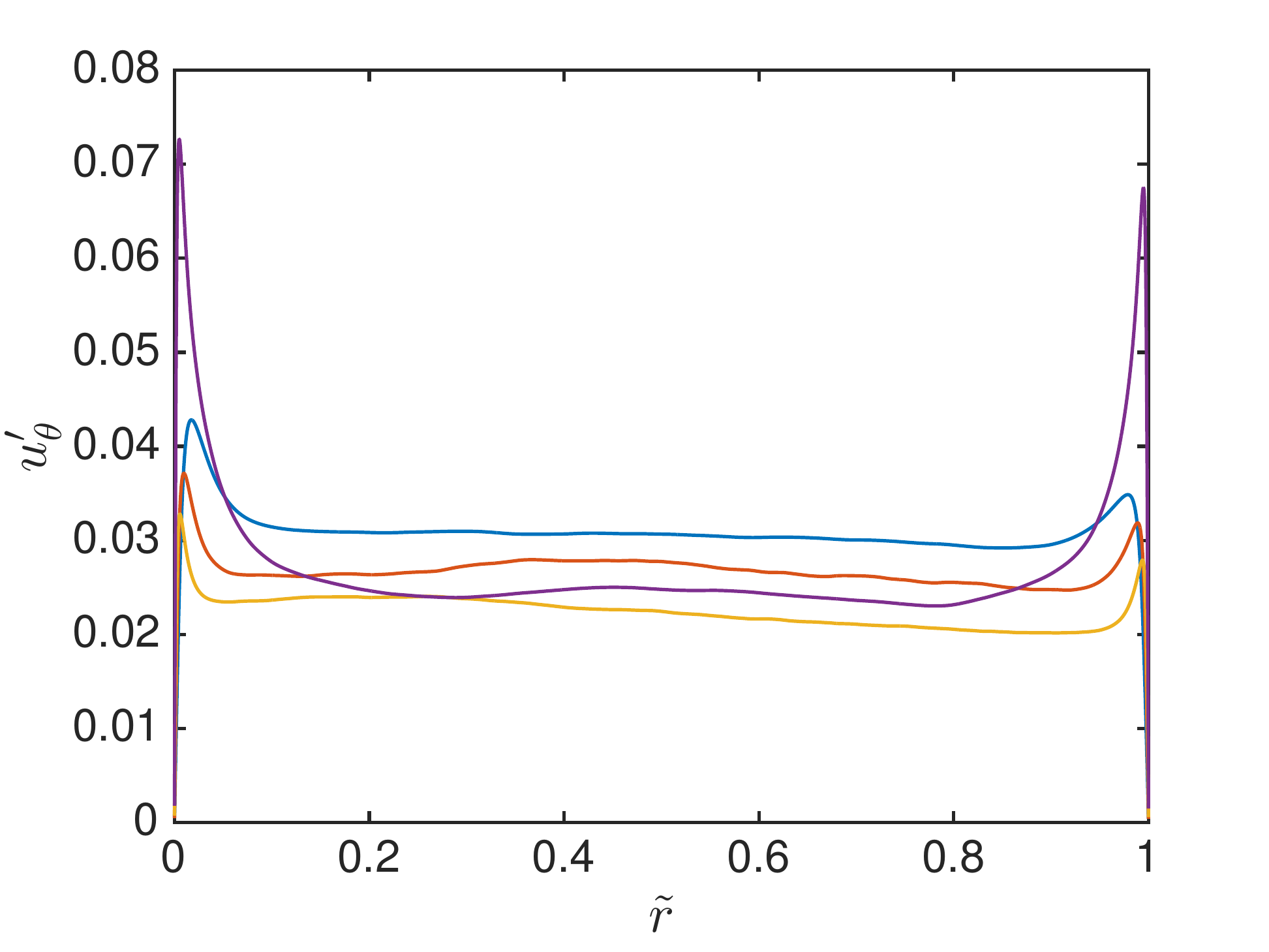}
     \caption{ The left panel shows the temporally, axially and azimuthally-averaged angular momentum for the three pure OCR cases and the I1 case for comparison, as well as experimental data from \cite{bur12}. The right panel shows the root mean square (r.m.s.) of the azimuthal velocity for the numerical cases. }
\label{fig:Louter}
\end{figure}

The left panel of figure \ref{fig:Louter} shows $\tilde{L}$, the azimuthally-, temporally- and axially- averaged angular momentum for the O5, O1, O2 and I1 cases, as well as the experimental data from \cite{bur12} for $\eta=0.97$ and $Re_o=6800$. For pure ICR rolls effectively redistribute angular momentum such that the flow is marginally stable. This is reflected in the I1 case showing a constant angular momentum profile in the bulk equal to the arithmetic average of $L$ at both cylinders. For pure OCR, the flow is already stable and thus we do not expect rolls to form and angular momentum not to be redistributed. Instead, all pure OCR cases show a significant gradient of angular momentum in the bulk. For pure OCR, the larger the angular momentum gradient in the bulk, the more stable the configuration. The resulting profile shape comes from the competing mechanisms of centrifugal stabilization in the bulk and destabilization in the boundary layers by shear. The numerical pure OCR velocity profiles are in qualitative agreement with the experimental profiles, as the bulk profile becomes more flat and the boundary layers thinner with increasing $Re_o$. We also note that similar phenomena were seen for the strongly counter-rotating cylinder cases of \cite{bra16}, which show significant deviations from constant angular momentum profiles in the bulk after the onset of the radial partitioning of stability.

The right panel of figure \ref{fig:Louter} shows root mean squared (r.m.s.) of the azimuthal velocity $u^\prime_\theta$ for the O5, O1, O2 and I1 cases. The level of fluctuations decreases with Reynolds number. When comparing the O1 and the I1 cases, a much lower level of fluctuations inside the boundary layer is seen for pure OCR. This is expected from the much lower values of $u_\tau$ for pure OCR, than for pure ICR. Indeed, $u_\tau$ is approximately a factor of two larger for the I1 case, and this is directly reflected in the level of fluctuations being approximately twice as large than the O1 case. 

\begin{figure}
  \centering
    \includegraphics[width=0.48\textwidth]{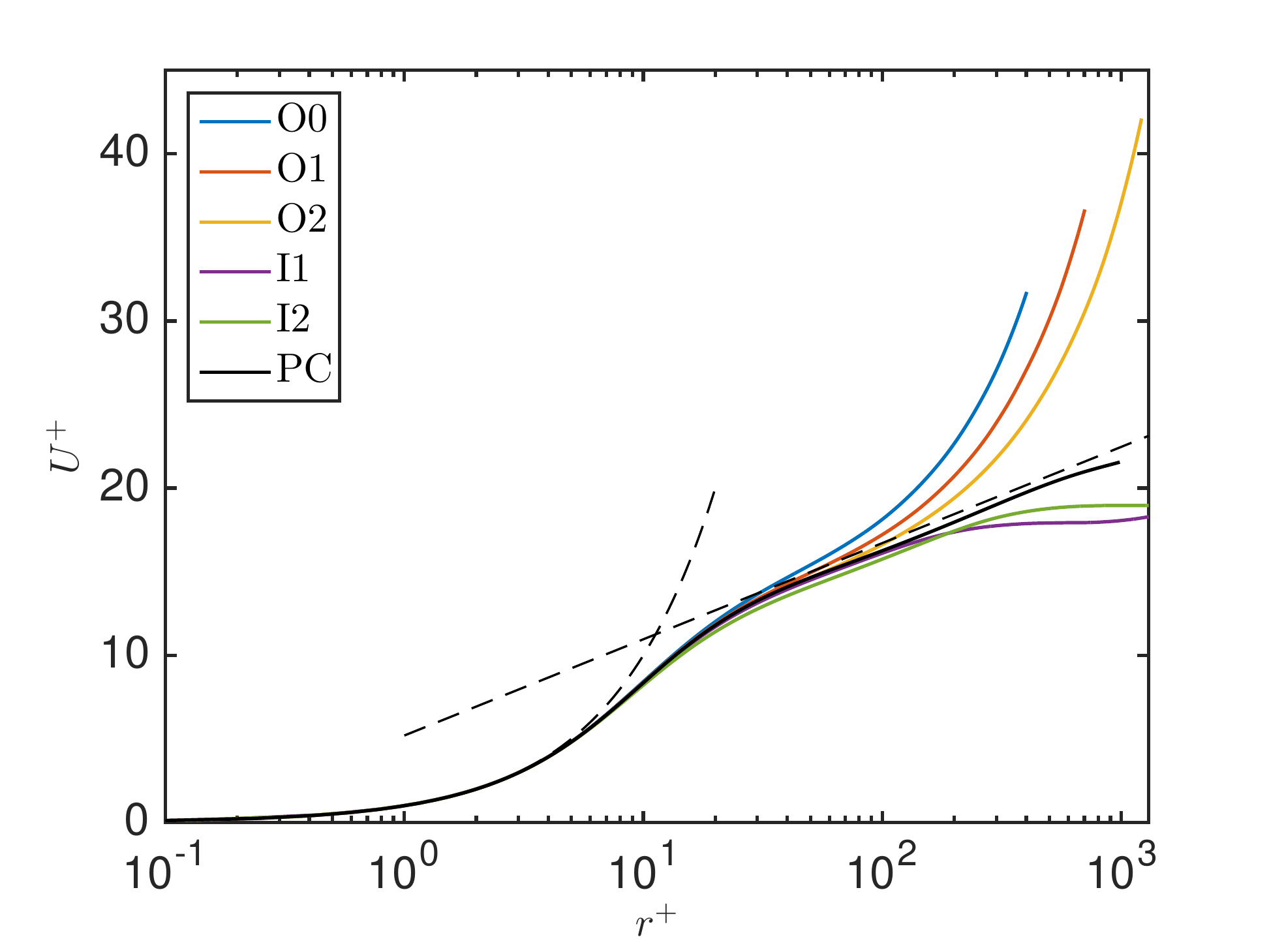}
     \includegraphics[width=0.48\textwidth]{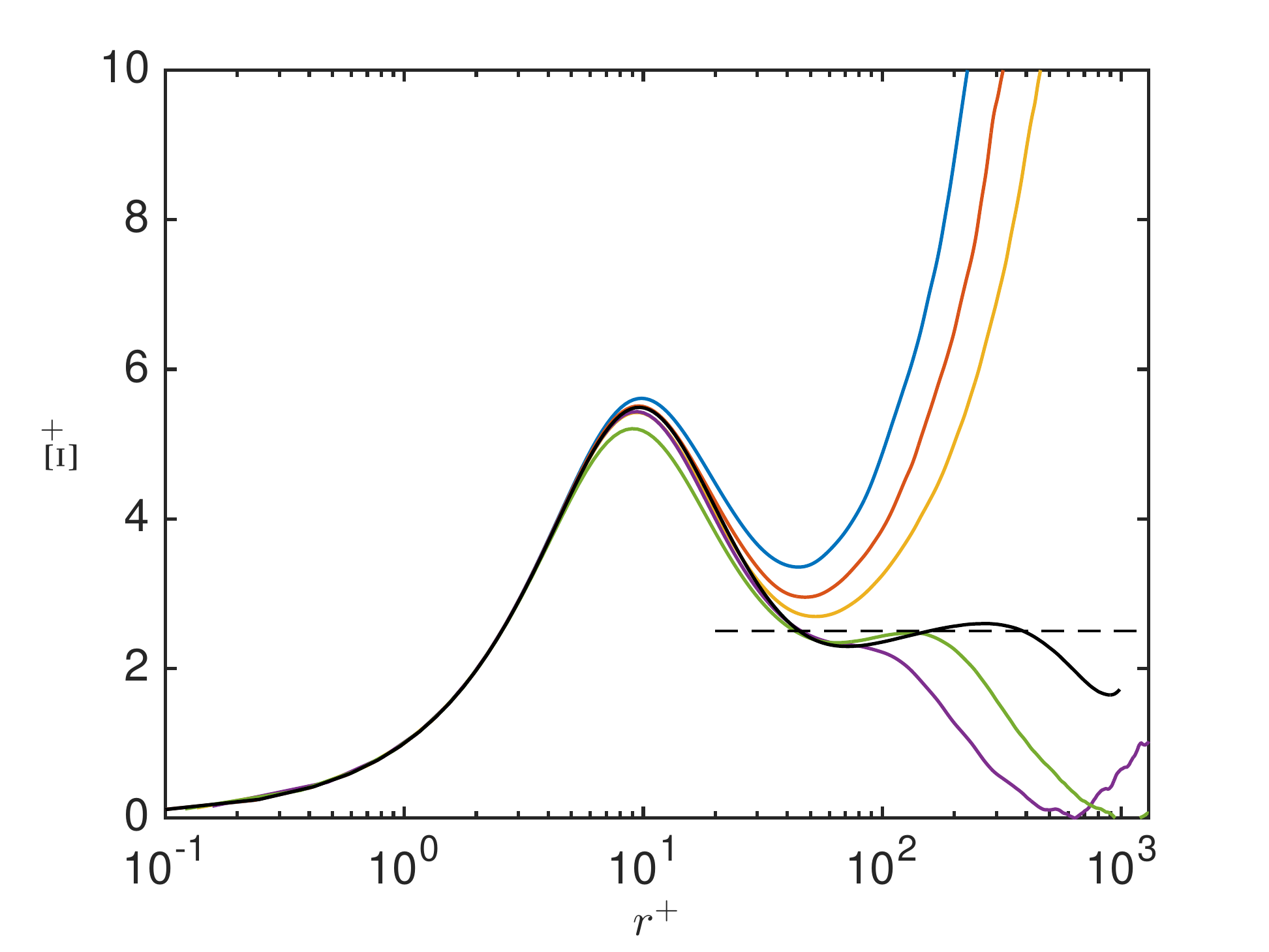}
     \caption{ The left panel shows the streamwise velocity in inner cylinder wall units for the all cases at the inner cylinder, and from PC flow at $Re_\tau\approx 1000$. The dashed curve represents $u^+=r^+$ and the dashed line represents $u^+=2.5\log(r^+)+5.2$. The right panel shows the logarithmic diagnostic function for the cases shown on the left panel. The horizontal dashed line represents $\Xi^+=2.5$.}
\label{fig:Uinner}
\end{figure}

We now focus on the near-wall region, to compare pure OCR with pure ICR, and to quantify the effects of curvature and instability. The left panel of figure \ref{fig:Uinner} shows the mean streamwise velocity at the inner cylinder region in inner units for all cases, where $U^+$ is $U^+=(r_i\omega_i - \langle u_\theta \rangle_{\theta,z,t})/u_{\tau,i}$ and $r^+$ is the distance from the wall in inner cylinder wall units $r^+=(r-r_i)/\delta_{\nu,i}$. The $Re_\tau\approx 1000$ plane Couette (PC) flow simulation from \cite{pir14} has been added for comparison.  Rotating PC flow is the limit of TC flow when $\eta\to 1$, i.e.~the two cylinders become two plates. Therefore, curvature effects and centrifugal (de)stabilization are not present. \cite{rom73} showed that PC flow is also stable to linear perturbations at all Reynolds numbers, though the mechanism is not centrifugal and this could cause different behavior.

Both pure ICR and pure OCR TC flow can be seen to deviate substantially from the classical von-Karman law of the wall $U^+=\kappa^{-1}\log(y^+) + B$, with $\kappa=0.4$ and $B=5.2$, while PC flow follows it better. Pure OCR TC flow has a significantly higher value of $U^+$ far away from the walls, while pure ICR TC flow has a rather flat profile in the bulk- consistent with the notion that angular momentum is redistributed in the bulk. It seems that while, in the bulk, pure ICR redistributes angular momentum through the rolls, pure OCR has the opposite effect, and generates a strong gradient of angular momentum. Thus, both lines deviate from the PC profile in opposite ways, showing the importance of the centrifugal (in)stability.
 
Very close to the wall, it could seem that the O2 case is beginning to show a logarithmic-like region. This can be better seen in the right panel of the figure, which shows the logarithmic diagnostic function $\Xi^+= d(\log(U^+))/dr^+$ for the same cases. Even if the pure OCR cases deviate much more than both the PC cases and the pure ICR cases, and do not show the S-like shape in $\Xi^+$ around $r^+\approx 100$ which is seen in several canonical flows \citep{ost16}, with increasing $Re_\tau$ number they are coming closer to the classical law-of-the-wall. It could be that for higher drivings, and thus higher $Re_\tau$ the pure OCR profiles collapse in the near-wal region on to the PC profiles, once the boundary layer does not feel the effect of curvature and of the centrifugal stabilization anymore. However, from the figures its seems that the centrifugal (in)stability mechanism plays a critical role in determining the bulk behavior, and is responsible for the large deviations of TC flow from PC flow behavior. Finally, the outer cylinder wall profiles show very similar behavior and are not shown here, so the main effects seems to be mediated by the centrifugal (in)stability and not by convex or concave curvature.

\begin{figure}
  \centering
    \includegraphics[width=0.48\textwidth]{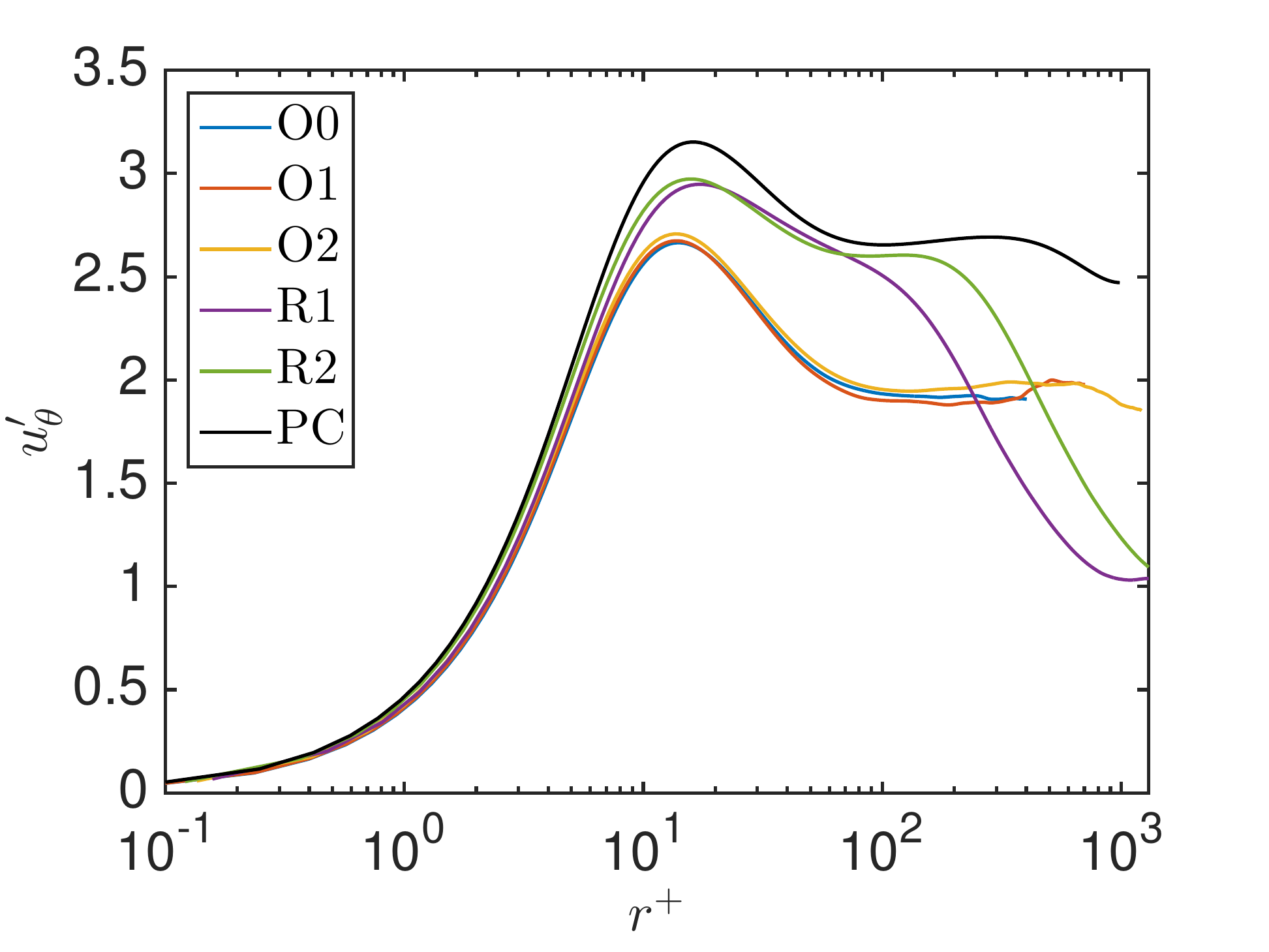}
     \includegraphics[width=0.48\textwidth]{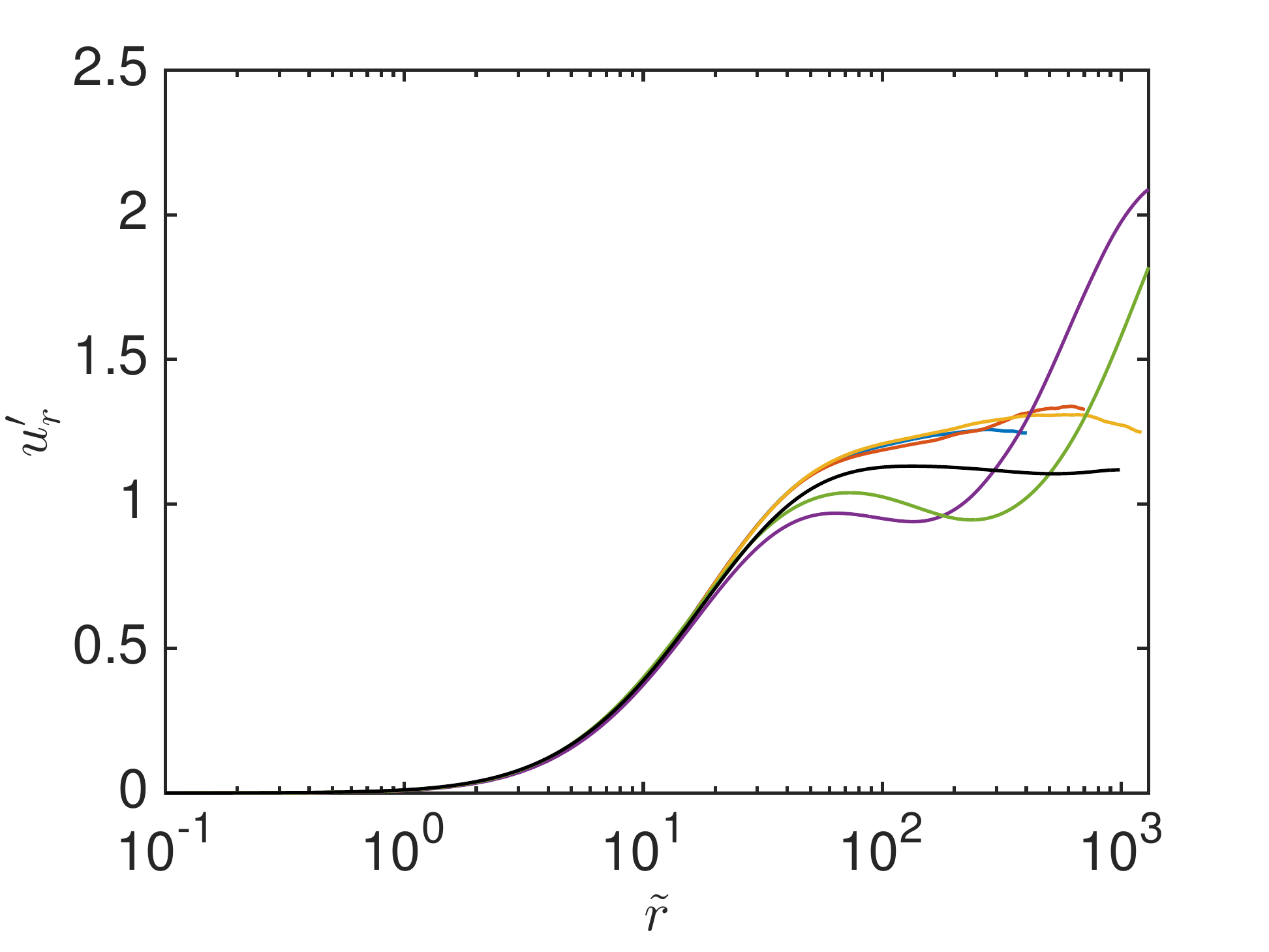}\\
      \includegraphics[width=0.48\textwidth]{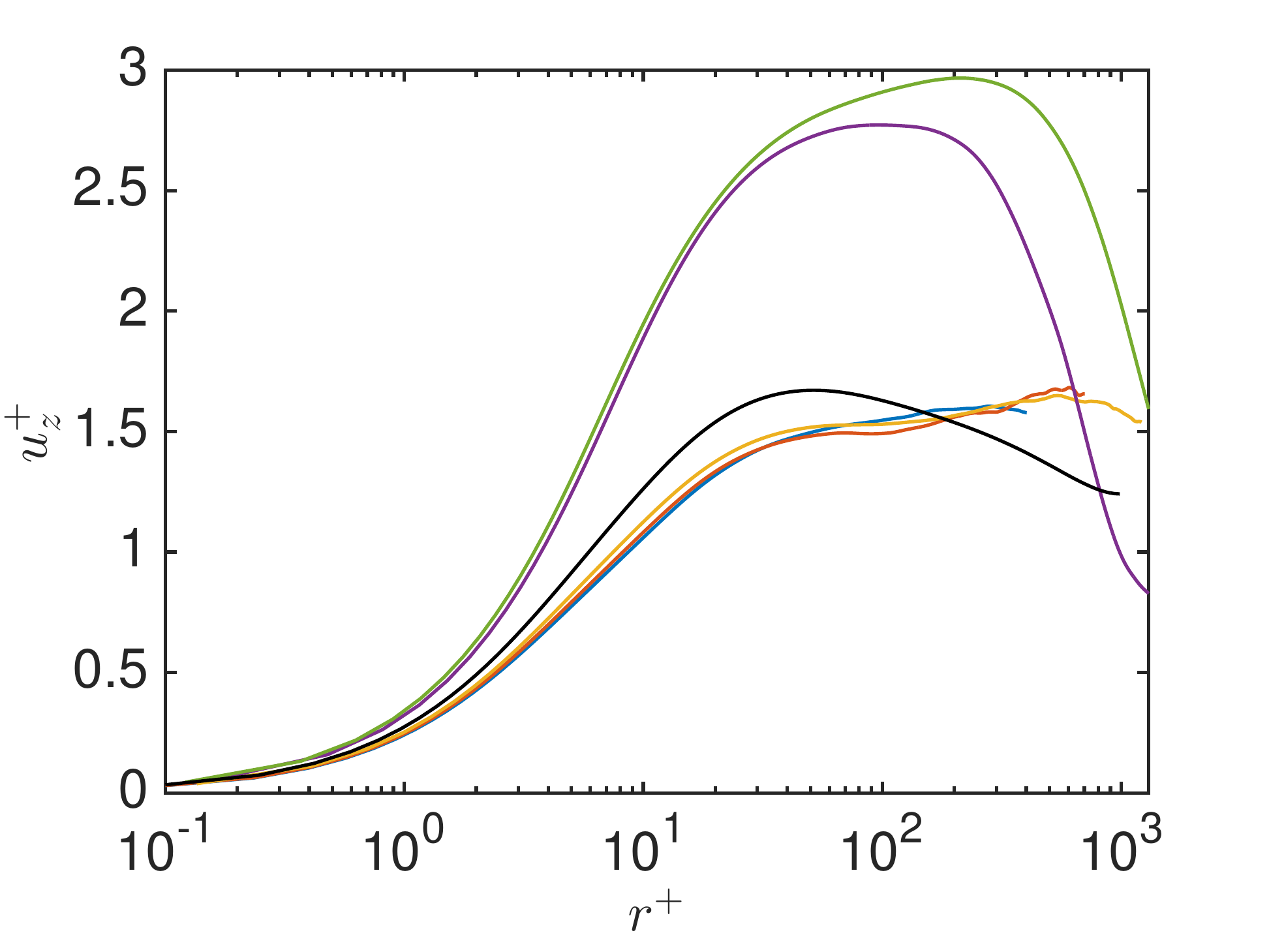}
     \caption{ Root-mean-squared fluctuations for all cases for the streamwise/azimuthal (top left), wall normal/radial (top right) and spanwise/axial (left) velocities in inner cylinder wall units.  }
\label{fig:primeinner}
\end{figure}

We now show the fluctuations in inner cylinder wall units for all three components of velocity in figure \ref{fig:primeinner}. While the streamwise fluctuations are considerably smaller for pure OCR, the profiles are closer to those of plane Couette flow for radial (wall normal) and axial (span-wise) fluctuations. The pure ICR rotation cases show very strong deviations for these two velocities, attributed to the axial inhomogeneity of the flow in \cite{ost16}. Again, the outer cylinder wall profiles show very similar behavior and are not shown here. The significant deviations from PC flow behavior can be attributed again to the different mechanisms at play, especially centrifugal (de)stabilization.

\begin{figure}
  \centering
    \includegraphics[width=0.48\textwidth]{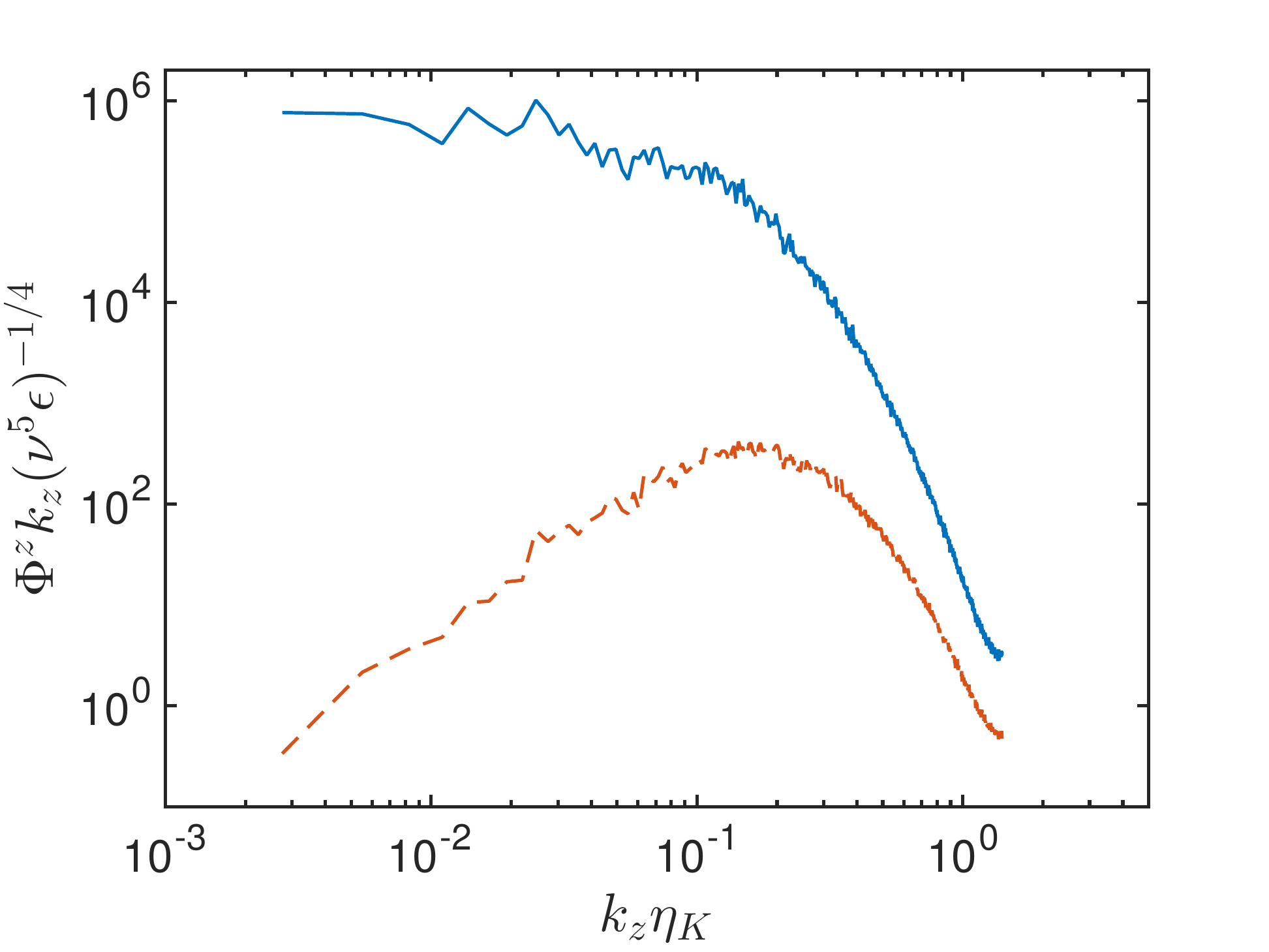}
    \includegraphics[width=0.48\textwidth]{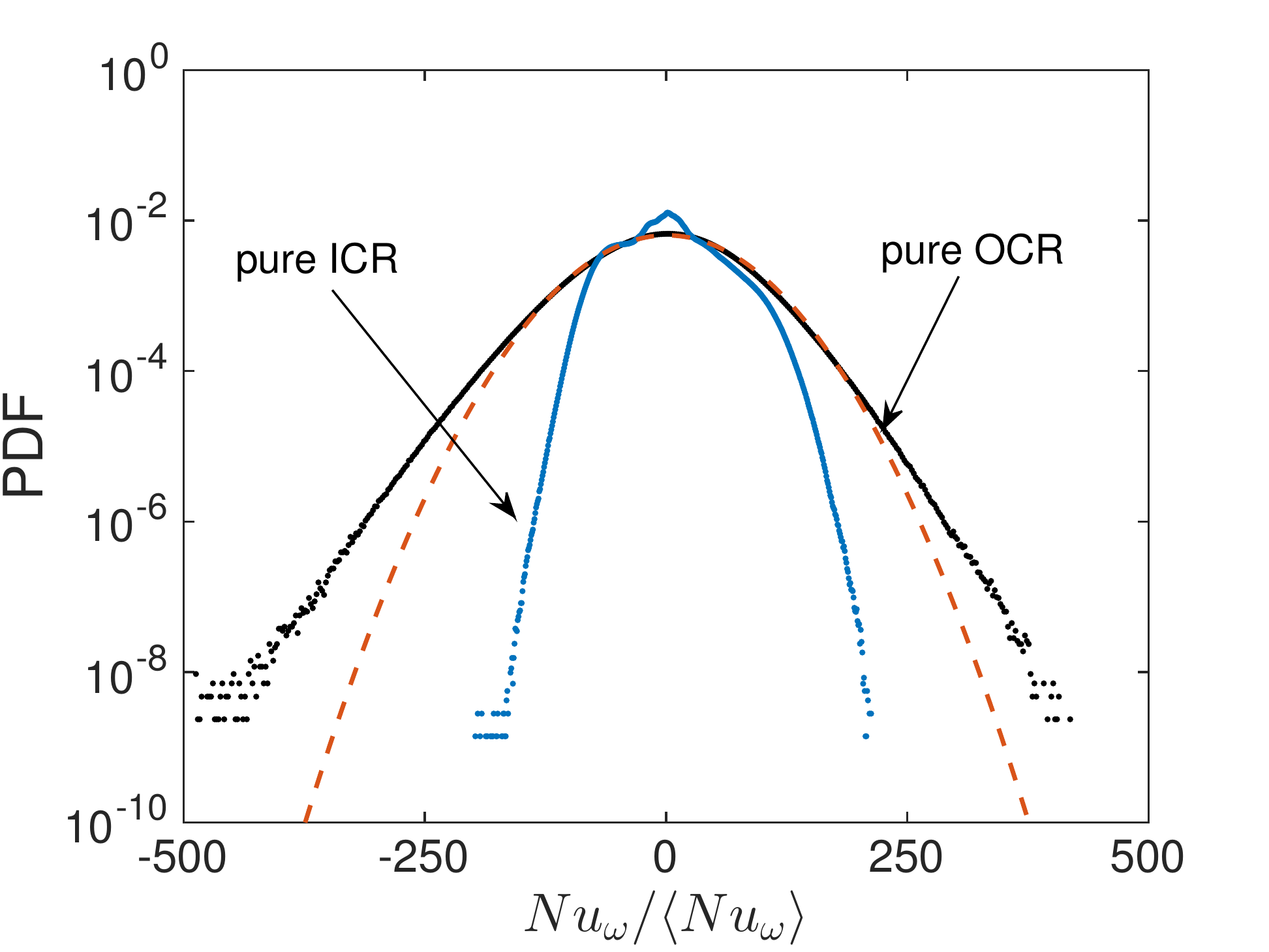}
     \caption{ The left panel shows the axial spectra for the radial (red dashed) and azimuthal (blue solid) velocities for $r^+\approx 12$, near the inner cylinder, for the O2 case. The right panel shows the p.d.f. of $Nu_\omega$ at the mid-gap for both the I2 and the O2 cases. The orange dash-dot line represents a Gaussian p.d.f. with mean and variance equal to that of the O2 case.}
\label{fig:spectra}
\end{figure}

Finally, to quantify the nature of transport in the boundary layers, the left panel of figure \ref{fig:spectra} shows the pre-multiplied axial spectra of radial and azimuthal velocity for the O2 case at $r^+\approx 12$, i.e.~around the peak of $u^\prime$ fluctuations inside the boundary layer. The peaks seen at the roll-wavelength for pure ICR in \cite{ost16} is no longer present,  as transport occurs through small scale fluctuations or 'bursts' \citep{bra13b}. These bursts transport angular velocity, and are very intermittent, having large amplitude but slow dynamics meaning that extreme events are more bound to happen. The peak in the radial spectra corresponds to the characteristic length-scale of these bursts. We note that spectra seen here are consistent with the spectra seen in channel flow \citep{jim12}, and in plane Couette flow \citep{avs14}, having a peak in the radial (wall-normal) spectra associated to the size of the transporting structures, and no saturation for the azimuthal velocity, indicating large-scale structures, attached to the wall which do not transport Reynolds stresses. 

To quantify this feature, the right panel of \ref{fig:spectra} shows the probability density function (p.d.f.) of the local convective angular velocity current $u_r\omega \approx Nu_\omega$ for both the O2 and the I2 case at mid-gap, i.e.\ in the bulk, as well as a Gaussian distribution with mean and variance equal to the O2 case. While for pure ICR, transport occurs mainly through the hairpin vortices, seen as the prominent peak centered around the middle of the graph, for the O2 case, the signature of this bursts is reflected here in the fatter tails of the p.d.f., which are super-Gaussian, and have no apparent power-law behavior. The p.d.f.s are not symmetric around zero, as there is a net positive angular velocity transport. From both panels, it becomes clear that the mechanisms for angular velocity transport are very different for pure ICR and pure OCR. 

\section{Summary and conclusions}

A series of DNS of turbulent Taylor-Couette flow with pure outer cylinder rotation were conducted. Overall, pure OCR TC flow behaves in a very different manner from supercritical pure ICR TC flow. The torque and fluctuation levels are much smaller for comparable Reynolds numbers than those of pure ICR flow. Transport of angular velocity, now more inefficient, occurs through intermittent ``bursts'', instead of through the large-scale structures. Pure OCR TC flow can be seen as just an extreme case of counter-rotating TC flow with the radial partitioning of stability described by \cite{bra13b,bra16} moving to the inner cylinder. The competition between the shear instabilities in the boundary layer and the centrifugal stabilization in the bulk gives rise to mean velocity profiles which show a significant angular momentum gradient in the bulk, consistent with the experiments of \cite{bur12}. The near-wall profiles deviate very strongly from both pure ICR rotation and plane Couette flow, revealing the very strong role of the centrifugal mechanisms in TC flow, be it stabilizing for pure OCR or destabilizing for pure ICR. Pure ICR and pure OCR deviate in opposite manners from the PC flow profiles, so this can be attributed to the role of centrifugal (de)stabilization. Finally, the large-scale structures completely disappear in this regime, and the axial velocity spectra reveal that transport near the wall occurs predominantly through very intermittent and small scale structures. 

Two main questions remain: the large-scale rolls seem to form in certain regions of the parameter space, where the flow is fully unstable \citep{ost14d}. However, it is still unclear why these rolls are formed, and why they are axially pinned. From these simulations, it seems that the centrifugal instability plays a clear roll in the nature of the turbulence and the formation of the rolls, but a complete understanding is still missing. Furthermore, the question on what happens in the quasi-Keplerian regime, which satisfies $|L_o| > |L_i|$, and $|\omega_o| < |\omega_i|$ remains \citep{ost14b,nor15}. These simulations have generated and sustained turbulence in the absence of end-plates at high Reynolds numbers. However, turbulence in the quasi-Keplerian regime has not been sustained in simulations, and, as mentioned previously this could be due to the opposing angular momentum and angular velocity gradients. We refer the reader to \cite{gro16} for a recent review of the progress on this problem. 

Acknowledgements: we thank V. Spandan for extensive help in proof-reading the paper and keeping some simulations running during the months-long wall-clock times, and we thank M. Burin for providing the data for figure \ref{fig:Louter} and for the valuable discussions. We acknowledge Y. Yang, and X. Zhu for fruitful and stimulating discussions. We also gratefully acknowledge computational time for the simulations provided by SurfSARA on resource Cartesius through a NWO grant. 

\bibliographystyle{jfm}
\bibliography{literatur}

\begin{thebibliography}{33}
\expandafter\ifx\csname natexlab\endcsname\relax\def\natexlab#1{#1}\fi
\def\au#1{#1} \def\ed#1{#1} \def\yr#1{#1}\def\at#1{#1}\def\jt#1{\textit{#1}}
  \def\bt#1{#1}\def\bvol#1{\textbf{#1}} \def\vol#1{#1} \def\pg#1{#1}
  \def\publ#1{#1}\def\arxiv#1{#1}\def\org#1{#1}\def\st#1{\textit{#1}}

\bibitem[Andereck {\em et~al.\/}(1983)Andereck, Dickman \& Swinney]{and83}
{\sc \au{Andereck, C.~D.}, \au{Dickman, R.} \& \au{Swinney, H.~L.}} \yr{1983}
  \at{New flows in a circular {{{Couette}}} system with corotating cylinders}.
  \jt{Phys. Fluids}  \bvol{26}~(1395).

\bibitem[Andereck {\em et~al.\/}(1986)Andereck, Liu \& Swinney]{and86}
{\sc \au{Andereck, C.~D.}, \au{Liu, S.~S.} \& \au{Swinney, H.~L.}} \yr{1986}
  \at{Flow regimes in a circular {{{Couette}}} system with independently
  rotating cylinders}.  \jt{J. Fluid Mech.}  \bvol{164},  \pg{155--183}.

\bibitem[van Atta(1966)]{vanatt66}
{\sc \au{van Atta, C.~W.}} \yr{1966}  \at{Exploratory measurements in spiral
  turbulence}.  \jt{J. Fluid Mech.}  \bvol{25}~(3),  \pg{495--512}.

\bibitem[Avsarkisov {\em et~al.\/}(2014)Avsarkisov, Hoyas, Oberlack \&
  {{{Garc\'{i}a-Galache}}}]{avs14}
{\sc \au{Avsarkisov, V.}, \au{Hoyas, S.}, \au{Oberlack, M.} \&
  \au{{{{Garc\'{i}a-Galache}}}, J.~P.}} \yr{2014}  \at{Turbulent plane
  {{{Couette}}} flow at moderately high {{{Reynolds}}} number}.  \jt{J. Fluid
  Mech.}  \bvol{751},  \pg{R1--8}.

\bibitem[Borrero-Echeverry(2014)]{bor14}
{\sc \au{Borrero-Echeverry, D.}} \yr{2014}  \at{Sub-critical transition to
  turbulence in {{{Taylor-Couette}}} flow}. PhD thesis, Georgia Institute of
  Technology, Atlanta, GA.

\bibitem[Borrero-Echeverry {\em et~al.\/}({2010})Borrero-Echeverry, Schatz \&
  Tagg]{bor10}
{\sc \au{Borrero-Echeverry, D.}, \au{Schatz, M.~F.} \& \au{Tagg, R.}}
  \yr{{2010}}  \at{{Transient turbulence in Taylor-Couette flow}}.  \jt{{Phys.
  Rev. E}}  \bvol{{81}},  \pg{025301}.

\bibitem[Brauckmann {\em et~al.\/}(2016)Brauckmann, Salewski \&
  Eckhardt]{bra16}
{\sc \au{Brauckmann, H.}, \au{Salewski, M.} \& \au{Eckhardt, B.}} \yr{2016}
  \at{Momentum transport in {{{Taylor-Couette}}} flow with vanishing
  curvature}.  \jt{J. Fluid Mech.}  \bvol{790},  \pg{419--452}.

\bibitem[Brauckmann \& Eckhardt(2013)]{bra13b}
{\sc \au{Brauckmann, H.~J.} \& \au{Eckhardt, B.}} \yr{2013}  \at{Intermittent
  boundary layers and torque maxima in {{{Taylor-Couette}}} flow}.  \jt{Phys.
  Rev. E}  \bvol{87}~(3),  \pg{033004}.

\bibitem[Burin \& Czarnocki(2012)]{bur12}
{\sc \au{Burin, M.~J.} \& \au{Czarnocki, C.~J.}} \yr{2012}  \at{Subcritical
  transition and spiral turbulence in circular {{{Couette}}} flow}.  \jt{J.
  Fluid Mech.}  \bvol{709},  \pg{106--122}.

\bibitem[Coles(1965)]{col65}
{\sc \au{Coles, D.}} \yr{1965}  \at{Transition in circular {{{Couette}}} flow}.
   \jt{J. Fluid Mech.}  \bvol{21},  \pg{385--425}.

\bibitem[Deguchi {\em et~al.\/}(2014)Deguchi, Meseguer \& Mellibovsky]{deg14}
{\sc \au{Deguchi, K.}, \au{Meseguer, A.} \& \au{Mellibovsky, F.}} \yr{2014}
  \at{Subcritical equilibria in {{{Taylor-Couette}}} flow}.  \jt{Phys. Rev.
  Lett.}  \bvol{112},  \pg{184502}.

\bibitem[Dubrulle {\em et~al.\/}(2005)Dubrulle, Dauchot, Daviaud, Longaretti,
  Richard \& Zahn]{dub05}
{\sc \au{Dubrulle, B.}, \au{Dauchot, O.}, \au{Daviaud, F.}, \au{Longaretti,
  P.~Y.}, \au{Richard, D.} \& \au{Zahn, J.~P.}} \yr{2005}  \at{Stability and
  turbulent transport in {{Taylor--Couette}} flow from analysis of experimental
  data}.  \jt{Phys. Fluids}  \bvol{17},  \pg{095103}.

\bibitem[Eckhardt {\em et~al.\/}(2008)Eckhardt, Faisst, Schmiegel \&
  Schneider]{eck08}
{\sc \au{Eckhardt, B.}, \au{Faisst, H.}, \au{Schmiegel, A.} \& \au{Schneider,
  T.}} \yr{2008}  \at{Dynamical systems and the transition to turbulence in
  linearly stable shear flows}.  \jt{Phil. Trans. R. Soc. A}  \bvol{366},
  \pg{1297--1315}.

\bibitem[Eckhardt {\em et~al.\/}(2007)Eckhardt, Grossmann \& Lohse]{eck07b}
{\sc \au{Eckhardt, B.}, \au{Grossmann, S.} \& \au{Lohse, D.}} \yr{2007}
  \at{Torque scaling in turbulent {{{Taylor-Couette}}} flow between
  independently rotating cylinders}.  \jt{J. Fluid Mech.}  \bvol{581},
  \pg{221--250}.

\bibitem[Fardin {\em et~al.\/}(2014)Fardin, Perge \& Taberlet]{far14}
{\sc \au{Fardin, M.~A.}, \au{Perge, C.} \& \au{Taberlet, N.}} \yr{2014}
  \at{The hydrogen atom of fluid dynamics - introduction to the
  {{{Taylor-Couette}}} flow for soft matter scientists}.  \jt{Soft Matter}
  \bvol{10}~(20),  \pg{3523--3535}.

\bibitem[van Gils {\em et~al.\/}(2012)van Gils, Huisman, Grossmann, Sun \&
  Lohse]{gil12}
{\sc \au{van Gils, D. P.~M.}, \au{Huisman, S.~G.}, \au{Grossmann, S.}, \au{Sun,
  C.} \& \au{Lohse, D.}} \yr{2012}  \at{Optimal {{Taylor-Couette}} turbulence}.
   \jt{J. Fluid Mech.}  \bvol{706},  \pg{118--149}.

\bibitem[Grossmann {\em et~al.\/}(2016)Grossmann, Lohse \& Sun]{gro16}
{\sc \au{Grossmann, S.}, \au{Lohse, D.} \& \au{Sun, C.}} \yr{2016}
  \at{{{{High--Reynolds}}} number {{{Taylor-Couette}}} turbulence}.  \jt{Ann.
  Rev. Fluid Mech.}  \bvol{48},  \pg{53--80}.

\bibitem[Huisman {\em et~al.\/}(2014)Huisman, van~der Veen, Sun \&
  Lohse]{hui14}
{\sc \au{Huisman, S.~G.}, \au{van~der Veen, R. C.~A.}, \au{Sun, C.} \&
  \au{Lohse, D.}} \yr{2014}  \at{Multiple states in highly turbulent
  {{{Taylor-Couette}}} flow}.  \jt{Nature Comm.}  \bvol{5},  \pg{3820}.

\bibitem[Jimenez(2012)]{jim12}
{\sc \au{Jimenez, J.}} \yr{2012}  \at{Cascades in wall-bounded turbulence}.
  \jt{Ann. Rev. Fluid. Mech.}  \bvol{44},  \pg{27--45}.

\bibitem[Nordsiek {\em et~al.\/}(2015)Nordsiek, Huisman, van~der Veen, Sun,
  Lohse \& Lathrop]{nor15}
{\sc \au{Nordsiek, F.}, \au{Huisman, S.~G.}, \au{van~der Veen, R. C.~A.},
  \au{Sun, C.}, \au{Lohse, D.} \& \au{Lathrop, D.~P.}} \yr{2015}  \at{Azimuthal
  velocity profiles in {{{Rayleigh}}}-stable {{{Taylor-Couette}}} flow and
  implied axial angular momentum transport}.  \jt{J. Fluid Mech.}  \bvol{774},
  \pg{342--362}.

\bibitem[Ostilla-Monico {\em et~al.\/}(2014{\natexlab{{\em
  a\/}}})Ostilla-Monico, van~der Poel, Verzicco, Grossmann \& Lohse]{ost14d}
{\sc \au{Ostilla-Monico, R.}, \au{van~der Poel, E.~P.}, \au{Verzicco, R.},
  \au{Grossmann, S.} \& \au{Lohse, D.}} \yr{2014{\natexlab{{\em a\/}}}}
  \at{Exploring the phase diagram of fully turbulent {{{Taylor-Couette}}}
  flow}.  \jt{J. Fluid Mech.}  \bvol{761},  \pg{1--26}.

\bibitem[Ostilla-Monico {\em et~al.\/}(2014{\natexlab{{\em
  b\/}}})Ostilla-Monico, Verzicco, Grossmann \& Lohse]{ost14b}
{\sc \au{Ostilla-Monico, R.}, \au{Verzicco, R.}, \au{Grossmann, S.} \&
  \au{Lohse, D.}} \yr{2014{\natexlab{{\em b\/}}}}  \at{Turbulence decay towards
  the linearly-stable regime of {{{Taylor-Couette}}} flow}.  \jt{J. Fluid
  Mech.}  \bvol{747},  \pg{1--29}.

\bibitem[Ostilla-M\'{o}nico {\em et~al.\/}(2016)Ostilla-M\'{o}nico, Verzicco,
  Grossmann \& Lohse]{ost16}
{\sc \au{Ostilla-M\'{o}nico, R.}, \au{Verzicco, R.}, \au{Grossmann, S.} \&
  \au{Lohse, D.}} \yr{2016}  \at{The near-wall region of highly turbulent
  {{{Taylor-Couette}}} flow}.  \jt{J. Fluid Mech.}  \bvol{768},  \pg{95--117}.

\bibitem[Ostilla-M\'{o}nico {\em et~al.\/}(2015)Ostilla-M\'{o}nico, Verzicco \&
  Lohse]{ost15}
{\sc \au{Ostilla-M\'{o}nico, R.}, \au{Verzicco, R.} \& \au{Lohse, D.}}
  \yr{2015}  \at{Effects of the computational domain size on {{{DNS}}} of
  {{{Taylor-Couette}}} turbulence with stationary outer cylinder}.  \jt{Phys.
  Fluids}  \bvol{27},  \pg{025110}.

\bibitem[Paoletti {\em et~al.\/}(2012)Paoletti, van Gils, Dubrulle, Sun, Lohse
  \& Lathrop]{pao12}
{\sc \au{Paoletti, M.~S.}, \au{van Gils, D. P.~M.}, \au{Dubrulle, B.}, \au{Sun,
  C.}, \au{Lohse, D.} \& \au{Lathrop, D.~P.}} \yr{2012}  \at{{{{Angular
  momentum transport and turbulence in laboratory models of Keplerian
  flows}}}}.  \jt{Astron. \& Astrophys}  \bvol{547},  \pg{A64}.

\bibitem[Paoletti \& Lathrop(2011)]{pao11}
{\sc \au{Paoletti, M.~S.} \& \au{Lathrop, D.~P.}} \yr{2011}  \at{Angular
  momentum transport in turbulent flow between independently rotating
  cylinders}.  \jt{Phys. Rev. Lett.}  \bvol{106},  \pg{024501}.

\bibitem[Pirozzoli {\em et~al.\/}(2014)Pirozzoli, Bernardini \& Orlandi]{pir14}
{\sc \au{Pirozzoli, S.}, \au{Bernardini, M.} \& \au{Orlandi, P.}} \yr{2014}
  \at{Turbulence statistics in {{{Couette}}} flow at high {{{Reynolds}}}
  number}.  \jt{J. Fluid Mech.}  \bvol{758},  \pg{327--343}.

\bibitem[van~der Poel {\em et~al.\/}(2015)van~der Poel, Ostilla-Monico, Donners
  \& Verzicco]{poe15}
{\sc \au{van~der Poel, E.~P.}, \au{Ostilla-Monico, R.}, \au{Donners, J.} \&
  \au{Verzicco, R.}} \yr{2015}  \at{A pencil distributed finite difference code
  for strongly turbulent wall-bounded flows}.  \jt{Comp. Fluids}  \bvol{116},
  \pg{10--16}.

\bibitem[{Rayleigh, Lord}(1917)]{ray17}
{\sc \au{{Rayleigh, Lord}}} \yr{1917}  \at{On the dynamics of revolving
  fluids}.  \jt{Proc. R. Soc. London A}  \bvol{93},  \pg{148--157}.

\bibitem[Romanov(1973)]{rom73}
{\sc \au{Romanov, V.~A.}} \yr{1973}  \at{Stability of plane-parallel couette
  flow}.  \jt{Functional analysis and its applications}  \bvol{7}~(2),
  \pg{137--146}.

\bibitem[Taylor(1923)]{tay23}
{\sc \au{Taylor, G.~I.}} \yr{1923}  \at{Experiments on the motion of solid
  bodies in rotating fluids}.  \jt{Proc. R. Soc. Lond. A}  \bvol{104},
  \pg{213--218}.

\bibitem[Taylor(1936)]{tay36}
{\sc \au{Taylor, G.~I.}} \yr{1936}  \at{Fluid friction between rotating
  cylinders}.  \jt{Proc. R. Soc. London A}  \bvol{157},  \pg{546--564}.

\bibitem[Verzicco \& Orlandi(1996)]{ver96}
{\sc \au{Verzicco, R.} \& \au{Orlandi, P.}} \yr{1996}  \at{A finite-difference
  scheme for three-dimensional incompressible flow in cylindrical coordinates}.
   \jt{J. Comput. Phys.}  \bvol{123},  \pg{402--413}.

\end{thebibliography}


\begin{thebibliography}{14}
\expandafter\ifx\csname natexlab\endcsname\relax\def\natexlab#1{#1}\fi

\bibitem[Batchelor(1971)]{Batchelor59}
{\sc Batchelor, G.~K.} 1971 Small-scale variation of convected quantities like
  temperature in turbulent fluid. part 1. general discussion and the case of
  small conductivity. {\em J.~Fluid Mech.\/} {\bf 5}, 113--133.

\bibitem[Brownell \& Su(2004)]{Brownell04}
{\sc Brownell, C.~J. \& Su, L.~K.} 2004 Planar measurements of differential
  diffusion in turbulent jets. {\em AIAA Paper 2004-2335\/}.

\bibitem[Brownell \& Su(2007)]{Brownell07}
{\sc Brownell, C.~J. \& Su, L.~K.} 2007 Scale relations and spatial spectra in
  a differentially diffusing jet. {\em AIAA Paper 2007-1314\/}.

\bibitem[Dennis(1985)]{Dennis85}
{\sc Dennis, S. C.~R.} 1985 {Compact explicit finite difference approximations
  to the Navier--Stokes equation}. In {\em Ninth Intl Conf. on Numerical
  Methods in Fluid Dynamics\/} (ed. Soubbaramayer \& J.~P. Boujot), {\em
  Lecture Notes in Physics\/}, vol. 218, pp. 23--51. Springer.

\bibitem[Hwang \& Tuck(1970)]{Hwang70}
{\sc Hwang, L.-S. \& Tuck, E.~O.} 1970 On the oscillations of harbours of
  arbitrary shape. {\em J.~Fluid Mech.\/} {\bf 42}, 447--464.

\bibitem[Koch(1983)]{Koch83}
{\sc Koch, W.} 1983 Resonant acoustic frequencies of flat plate cascades. {\em
  J.~Sound Vib.\/} {\bf 88}, 233--242.

\bibitem[Lee(1971)]{Lee71}
{\sc Lee, J.-J.} 1971 Wave-induced oscillations in harbours of arbitrary
  geometry. {\em J.~Fluid Mech.\/} {\bf 45}, 375--394.

\bibitem[Linton \& Evans(1992)]{Linton92}
{\sc Linton, C.~M. \& Evans, D.~V.} 1992 The radiation and scattering of
  surface waves by a vertical circular cylinder in a channel. {\em Phil.\
  Trans.\ R. Soc.\ Lond.\/} {\bf 338}, 325--357.

\bibitem[Martin(1980)]{Martin80}
{\sc Martin, P.~A.} 1980 On the null-field equations for the exterior problems
  of acoustics. {\em Q.~J. Mech.\ Appl.\ Maths\/} {\bf 33}, 385--396.

\bibitem[Miller(1991)]{Miller91}
{\sc Miller, P.~L.} 1991 Mixing in high schmidt number turbulent jets. PhD
  thesis, California Institute of Technology.

\bibitem[Rogallo(1981)]{Rogallo81}
{\sc Rogallo, R.~S.} 1981 Numerical experiments in homogeneous turbulence. {\em
  Tech. Rep.\/} 81835. NASA Tech.\ Mem.

\bibitem[Ursell(1950)]{Ursell50}
{\sc Ursell, F.} 1950 Surface waves on deep water in the presence of a
  submerged cylinder i. {\em Proc.\ Camb.\ Phil.\ Soc.\/} {\bf 46}, 141--152.

\bibitem[{van Wijngaarden}(1968)]{Wijngaarden68}
{\sc {van Wijngaarden}, L.} 1968 On the oscillations near and at resonance in
  open pipes. {\em J.~Engng Maths\/} {\bf 2}, 225--240.

\bibitem[Worster(1992)]{Worster92}
{\sc Worster, M.~G.} 1992 {The dynamics of mushy layers}. In {\em In
  Interactive dynamics of convection and solidification\/} (ed. S.~H. Davis,
  H.~E. Huppert, W.~Muller \& M.~G. Worster), pp. 113--138. Kluwer.

\end{thebibliography}

\end{document}